# *Digital Metamaterials*

*Cristian Della Giovampaola and Nader Engheta*[*]

*University of Pennsylvania*
*Department of Electrical and Systems Engineering*
*Philadelphia, Pennsylvania 19104, USA*
Email: dcr@seas.upenn.edu, engheta@ee.upenn.edu

## *Abstract*

**Balancing the complexity and the simplicity has played an important role in the development of many fields in science and engineering. As Albert Einstein was once quoted to say: "*Everything must be made as simple as possible, but not one bit simpler*"[1,2]. The simplicity of an idea brings versatility of that idea into a broader domain, while its complexity describes the foundation upon which the idea stands. One of the well-known and powerful examples of such balance is in the Boolean algebra and its impact on the birth of digital electronics and digital information age[3,4]. The simplicity of using only two numbers of "0" and "1" in describing an arbitrary quantity made the fields of digital electronics and digital signal processing powerful and ubiquitous. Here, inspired by the simplicity of digital electrical systems we propose to apply an analogous idea to the field of metamaterials, namely, to develop the notion of *digital metamaterials*. Specifically, we investigate how one can synthesize an electromagnetic metamaterial with desired materials parameters, e.g., with a desired permittivity, using only two elemental materials, which we call "*metamaterial bits*" with two distinct permittivity functions, as building blocks. We demonstrate, analytically and numerically, how proper spatial mixtures of such metamaterial bits leads to "*metamaterial bytes*" with material parameters different from the parameters of metamaterial bits. We also explore the role of relative spatial orders of such digital materials bits in constructing different parameters for the digital material bytes. We then apply this methodology to several design examples such as flat graded-index digital lens, cylindrical scatterers, digital constructs for epsilon-near-zero (ENZ) supercoupling[5–8], and digital hyperlens[9–11], highlighting the power and simplicity of this methodology and algorithm**.

---
[*] To whom correspondence should be addressed, Email: engheta@ee.upenn.edu

## *Main Text*

Electromagnetic and optical fields and waves may be manipulated and controlled by material media in various wave-based devices and systems. Numerous materials with a wide range of electromagnetic parameters have been utilized to achieve this task. In addition to naturally available materials, in recent years the concept of artificially engineered materials, known as metamaterials, have provided access to even a broader range of parameter space, such as negative, near-zero, or high values for constituent material parameters of permittivity, permeability, refractive index, chirality, and nonlinear susceptibility, to name a few[12–16]. In a variety of scenarios and for various reasons, one may not always have access to all materials with all the required parameter values. For example, materials with extremely high dielectric constant in the visible and near infrared (NIR) are hard to find. A question may then arise: What if the choice of materials is limited to only a few, e.g., only two materials, in a given range of operating wavelengths? May we still obtain the desired parameter values by properly combining and sculpting the two given constituent materials with different parameters, particularly when the desired parameter values are vastly different from those of the constituent materials? Indeed, this quest is consistent with the notion of metamaterials, and many examples of metamaterials constructed in the past decade by various research groups worldwide have been built using a limited number of materials[17–24]. However, here we develop the notion of *digital metamaterials* in order to demonstrate that with only two properly chosen elemental materials, coined as "metamaterial bits", at a given range of operating wavelengths, one would, under proper conditions, be able to synthesize composite media with a large range of parameter values. We also offer specific quantitative recipes of how to combine the two metamaterial bits at each point in order to obtain the desired parameter values. This simple, yet powerful and ubiquitous, concept and recipe may simplify the design of devices and systems, may provide parameter values in a wavelength regime in which natural materials with such desired values may not be available, and may facilitate and ease nanofabrication of composites with more complex properties. Our vision, inspired by the Boolean algebra and the digital electrical system, is summarized in Fig. 1. In digital electronics (Fig. 1a), to represent an analog function of time in terms of its digital bits (downward arrow), one first samples the analog signal (sampling process), then the sampled data are discretized (quantization process) and every discretized number at a given sample point is converted to a digital signal using a binary algebra (analog-to-digital conversion). Such digital signals are then processed with digital logics based on the Boolean algebra. In the reverse processes (upward arrow), the digital bytes can be converted to analog signals using the standard digital-to-analog conversion, recovering the discretized analog signals that need to go through the low-pass filter in order to provide a signal similar to the original signal. For digital metamaterials (Fig. 1b), we propose an analogous path: when a desired permittivity as a function of spatial coordinates is sought (in Fig. 1b, without loss of generality we assume a two-dimensional $x$-$y$ system), this function should also first be spatially sampled and discretized, and then for every sample point, or cell, in the $x$-$y$ plane, we should design a proper combination of two elemental "bit" materials (shown as the gray and green elements in Fig. 1b) such that the effective permittivity of this properly mixed combination in a cell (shown as a "jigsaw puzzle" mixture) would become the discretized value of desired permittivity for that cell. In the binary algebra, the properly ordered sequence of "0's" and "1's" determines the proper digital number for the sampled signal value at each cell. In the digital metamaterial paradigm, as will be shown later, the spatial ordering of metamaterial bits also determines the effective permittivity for that cell. The choice of topology of the mixture is guided by the style suitable for nanofabrication. We consider two different styles of sculpting the metamaterial bits: (1) planar or rectangular bits, and (2) core-shell or concentric bits. In the first style, the bits are in a rectangular form as shown in the right inset given in the bottom of Fig. 1b, whereas for the second style we consider a concentric pair of bits in the form of cylindrical or spherical core-shell (left inset in the bottom of Fig. 1b). The rectangular bits can also provide some degree of anisotropy and polarization dependence in the call as we will show later in this work, while the core-shell bits lead to approximately isotropic material cells (in the x-y plane). One of the interesting advantages of digitizing the material systems is that the two material bits can be selected to have favorable characteristic within the range of frequency of interest, e.g., low-loss bits, flexible bits, reconfigurable bits, etc.

*Metamaterial bits and bytes*: Top row in Figs. 2a and 2b shows the schematic of our two configurations of metamaterial bytes formed by 2-dimensional (2D) mixtures of two material bits, depicted as the green and gray regions. For the rest of our discussion, without loss of generality only two-dimensional scenarios in which all quantities are independent of the $z$ coordinate are considered for the squared-cross-section (Fig. 2a) and concentric-core-shell (Fig. 2b) cylindrical cells. We note that these two cell configurations are only examples under consideration here for the sake of mathematical simplicity in the analysis of wave interaction with them; however, other geometrical configurations and arrangements for material compositions may also be utilized. We assume that the cross-sectional size of the 2D cell is deeply subwavelength, which provides two advantages in our design: (1) As in digital electronics, here the smaller cell size provides discrete (stepwise) effective permittivity distributions that resemble more closely to the desired smooth permittivity function; and (2) the small cross-sectional size allows us to utilize simplifying approximation for electromagnetic wave interaction with such material bytes (e.g., quasi-static small dimension approximation). In each byte, the two material bits occupy certain volume fractions of the cell. To gain physical intuition on the effects of the geometry and spatial ordering of the material bits in each case, let us consider the squared-cross-section configuration in Fig. 2a for two different orientations as shown, and let us utilize the simplified analytical expressions for the effective permittivity of this 2D cell[22,25], expressed as $e_{\parallel} = \dfrac{n\,e_m + (N-n)\,e_d}{N}$   ;   $e_{\wedge} = \dfrac{N\,e_m\,e_d}{(N-n)\,e_m + n\,e_d}$ where the subscripts $\parallel$ and $\wedge$ refer, respectively, to the cases of parallel and perpendicular polarization of the incident wave with respect to the boundary between the two material bits with relative permittivity $e_m$ and $e_d$ in the cell, and $\dfrac{n}{N}$ and $1 - \dfrac{n}{N}$ are the volume fraction of materials bits, respectively, as shown in Fig. 2a. A simple examination of these expressions reveals that $e_{\parallel}$ attains values that are effectively arithmetic weighted average of the permittivity functions of the two material bits. However, the effective parameter $e_{\wedge}$ may possess values outside the range of values between $e_m$ and $e_d$, provided these two permittivity functions have oppositely signed real parts. For example, one of the bits can be a dielectric material with positive permittivity, while the other may be a metal with negative permittivity at optical and infrared (IR) wavelengths (hence the choice of subscripts "$d$" and "$m$"). The plots in Fig. 2a show the real and imaginary parts of the relative effective permittivity $e_{\parallel}$ (left panel in Fig. 2a) and $e_{\wedge}$ (right panel in Fig. 2a) as a function of volume fraction of $e_m$, assuming $e_d$ a real quantity and $e_m = e'_m + i e''_m$ a complex quantity. We clearly see that with proper combination and orientation of the two material bits, we can achieve a large range of possible values for the effective permittivity of the squared-cross-section cell, as our metamaterial byte. As expected, the wide range of available values for the effective permittivity of our cell is ultimately limited by the loss in $\varepsilon_m$. Nevertheless this range is far larger than, and outside the range of, values of the two material bits. This shows that with the two material bits, one can design a material byte with the effective permittivity vastly different from the permittivity values of the two bits. An analogous analysis may be conducted for the concentric core-shell cylindrical byte shown in Fig. 2b. When this 2D structure is illuminated by an electromagnetic plane wave with transverse magnetic (TM) polarization (i.e., the electric field lies in the *x-y* plane that is orthogonal to the cylinder axis along the *z* direction), its effective permittivity for the small-radii approximation, following the so-called 'internal homogenization'[26] procedure, can be written as $\varepsilon_{cs} = \varepsilon_s \dfrac{\varepsilon_c + \varepsilon_s + r_{cs}^2 \left( \varepsilon_c - \varepsilon_s \right)}{\varepsilon_c + \varepsilon_s - r_{cs}^2 \left( \varepsilon_c - \varepsilon_s \right)}$ where $\varepsilon_c$ and $\varepsilon_s$ are the relative permittivity of the core and of the shell, respectively, and $r_{cs} = a/b$, with *a* and *b* being the radius of the core and the shell, respectively, as shown in Fig. 2b. As in the squared-cross-section byte, here the core-shell cell also exhibits the possibility of accessing a large range of values for the relative effective permittivity $\varepsilon_{cs}$ by properly selecting the ratio of radii and the proper order of material bits for the core and shell. It is interesting to note that, analogous to the Boolean binary algebra where the order of the bits affect the value of the digital byte, here the order of the material bits with the positive and negative permittivity for the core and the shell also significantly affects the value of the effective permittivity of the material byte, as clearly sketched in Fig. 2b. Before we proceed to demonstrate the power of this concept, it is worth highlighting the fact that the choice of the two material bits is in general arbitrary, so long as the real parts of their permittivity functions are oppositely signed in the wavelength range of interest. Other considerations, such as mechanical conditions, low conductive loss, thermal properties, and nanofabrication ease may be employed in narrowing down this choice. Moreover, as we will show in the rest of the present work, the same two material bits can be utilized to design various different structures and devices, emphasizing the power of the simplicity of this concept, and its analogy to the "1's" and "0" bits in digital electronics. As an example, we choose silver (Ag) and silica (SiO$_2$) to be our two material bits, operating at the wavelength of 405 nm (of a GaN laser which is used, for example, in regular blue-ray disc readers), thus exhibiting the relative permittivity values of[27] $\varepsilon_m = -4.70 + i0.22$ and[28] $\varepsilon_d = 2.42$, respectively. With the above selection of materials, the effective permittivity for the core-shell cell follows the behavior shown in Fig. 2b. It is worth noting that the achievable range of permittivity (in the presence of loss) depends on the relative value of $\varepsilon_d$ and $\varepsilon_m$, as can be determined from the above internal homogenization equation.

Through a series of numerical simulations, we now demonstrate the utility (and also some of the limitations) of the concept of digital metamaterials in design of several structures with special functionalities.

*Scattering characteristics*: We first evaluate the scattering of electromagnetic waves from our two configurations of digital metamaterial bytes, and then compare the results with the scattering from the analogous structures possessing their equivalent effective permittivity values. This comparison reveals how closely a digital metamaterial byte resembles, and indeed imitates the electromagnetic properties of, its equivalent cell with the desired effective permittivity, when it is interrogated by electromagnetic waves from outside. As shown in Figs. 2c-2f, this scattering comparison is performed on infinitely extent cylinders made of different configurations of our material bits and for different polarizations of incident wave. The layered bytes are shown in Figs. 2c-2d, while the concentric core-shell case is presented in Figs. 2e-2f. For the layered cases of paired Ag-SiO$_2$, we consider cylinders with radius $\lambda_0/5$ where $\lambda_0$ is the free space wavelength, while the thickness of each Ag-SiO$_2$ pair is $\lambda_0/20$ with the thickness fraction of Ag to be 0.6 in Fig. 2c and 0.309 in Fig. 2d. The transverse magnetic (TM in Fig. 2c) and transverse electric (TE in Fig. 2d) polarizations for the incident wave are considered. According to the approximate analytical formula given above, the effective permittivity of the metamaterial byte in Fig. 2c is $-1.85+i0.13$, while that of Fig. 2d is $4.43+i0.05$. We note that in the former case, the effective value is between the two permittivity values of Ag and SiO$_2$, while for the latter, the effective value is outside the range. The results of our numerical simulations on the digital metamaterial bytes shown in Figs. 2c-2d and the corresponding 2D infinite cylinders of the same outer radius, but with the homogenous permittivity with the effective values are shown in the Figure. The color maps (with the same logarithmic scale bars) show the distributions of the magnitude of the scattered electric fields (in V/m in dB, with the amplitude of the incident being 0 dBV/m) for each pair of digital byte and its corresponding "homogenous" cylinder. The excellent agreement between the scattered fields distributions for the digital case and its homogenous counterpart demonstrate the notion that, as viewed from the outside, indeed the two-bit metamaterial mixtures behave similar to their homogeneous cases with the effective parameter values. Needless to say the field distributions inside the structures are vastly different in the "digital" and the "homogeneous" cases, while the scattered fields observed outside are similar. In a similar fashion, the concentric-core-shell configuration has also been studied and reported in Figs. 2e-2f. Here the cylinder has radius of $\lambda_0/40$, but the order of Ag and SiO$_2$ is different in Fig. 2e (SiO$_2$ as core and Ag as shell and radii ratio of 0.4) and Fig. 2f (SiO$_2$ as shell and Ag as core, the radii ratio being 0.48), both with the TM incident wave. The values of the effective permittivity of these two cases are $-1.57 + i0.21$ and $14.12 + i2.71$, respectively. (We note again that the former value is between the two values of Ag and SiO₂, while the latter is significantly outside this range.) The scattering comparison again reveals that these digital bytes behave similarly to their corresponding homogenous cylinders. Here, we reiterate that by using two materials (Ag and SiO₂) with permittivity values of $\varepsilon_m = -4.70 + i0.22$ and $\varepsilon_d = 2.42$, one can design a cylinder with a high effective permittivity value of $14.12 + i2.71$, highlighting the power of digital metamaterials. Using the digital bytes shown in Fig. 2, we now design several structures with various functionalities.

*Digital convex lens*: The first example, shown in Fig. 3a, is the digital version of a 2D dielectric convex lens with hyperbolic profile. The lens width is $6\lambda_0$, its maximum thickness in its middle section is approximately $1.2\lambda_0$ and the focal line is located $1.8\lambda_0$ away from the lens edge. With these dimensions and focal distance, the required relative permittivity for the lens is $e_{DL} = 4$. Can we achieve such relative permittivity using only two material bits of Ag and SiO₂ with the use of one of our digital metamaterial bytes shown in Fig. 2? This is indeed possible as shown here: The region corresponding to the 2D lens may be filled with many identical core-shell digital cylindrical bytes, each with diameter of $\lambda_0 / 20$, and a proper ratio of core-shell radii of 0.542 in order to provide the desired effective permittivity. In the arrangement we used in this simulation, as well as in the following examples, the core-shell cells do not touch each other, but some air gap is intentionally left between them in order to avoid possible hotspots for the electric field. The effect of such air gaps has been taken into account in our effective parameter numerical retrieval. Since the desired effective permittivity $e_{DL}$ is greater than $e_d$, according to the expression for the effective permittivity of the core-shell byte, the core has to be made of Ag while the shell should be made of SiO2. Such a digital lens (shown in Fig. 3a) is then numerically simulated when it is illuminated by a plane wave impinging normally on the flat part of the lens, with the electric field vector in the *xy* plane (TM polarization, similar to Fig. 2f configuration). The field map of Fig. 3a shows the distribution of the amplitude of the electric field vector, demonstrating the field concentration around the desired focal line, and therefore confirming its proper performance, which resembles the performance of a lens with the desired effective permittivity $\varepsilon_{DL}$. Moreover, although we take into account the loss of Ag in the complex nature of its permittivity and consequently the core-shell byte also possesses loss, this dissipation does not cause a substantial reduction of the field intensity at the focal line of such digital lens (for results concerning the effects of the losses on the lens performance, see the Supplementary Information).

*Digital graded-index flat lens*:  Since our digital metamaterial bytes have deeply subwavelength cross sections, they can be used as building blocks for constructing inhomogeneous distributions of effective permittivity in various structures. In Figs. 3b and 3c, we present two different designs for 2D digital graded-index flat lenses with rectangular cross section, requiring inhomogeneous spatial distributions of their effective permittivity. The dimensions of the lens shown in Figs. 3b (the same as for the lens in Fig. 3c) determine the required variation of relative permittivity along the transverse direction (the spatial variation of the permittivity used to design the lens is shown in the Supplementary Information). Here the required relative permittivity for both lenses ranges from 4 (at the center of the slab) to 1 (at the edges). We note that some of these required values to be synthesized by our digital bytes lie in between the two permittivity values of our material bits (Ag and SiO$_2$), while the other values lie outside this interval. As discussed above, such range of value can indeed be constructed by our properly designed digital bytes. The two digital lenses shown in Figs. 3b-3c are designed with two different digital bytes, core-shell bytes (Fig. 3b) and layered bytes (Fig. 3c), for two different polarizations of the source field. However, both possess the same graded distribution for the effective relative permittivity, therefore demonstrating that the underlying concept and mechanism behind the digital metamaterials is independent of the specific arrangement of the two material bits, provided each configuration is designed such that it would cover the required range of effective permittivity. By following the strategy described in the previous examples, when considering the core-shell digital bytes (Fig. 3b), the bytes in the central region of the lens (where higher effective permittivity is required) need to have a Ag core and a SiO$_2$ shell, while the ones towards the side regions should have a silica core and a Ag shell. Similarly, for the lens made of layered rectangular bytes (Fig. 3c), the central elements have the layers orthogonal to the polarization vector of the incident electric field, whereas on the sides the layers are parallel to it. The transition between the two configurations happens when the required effective permittivity function reaches the value of permittivity of SiO$_2$ (this is not exactly the case for the core-shell bytes due to the presence of air gaps between cylinders). In the core-shell digital lens of Fig. 3b, the source is an infinitely long line of magnetic current located at and parallel with the focal line. This radiates an electric field vector in the *xy* plane, suitable for the core-shell bytes lens. The field map in Fig. 3b presents the proper operation of such digital lens, showing a planar wavefront leaving the lens. As an aside, the effect of the material loss in Ag does not appreciably affect the lens performance since the field attenuation through the lens is limited. In Fig. 3c we report the simulation results for the digital lens made of squared-layered bytes, with the arrangement of the layers shown. Unlike the previous case, here the source is an infinitely long line of electric current, again at the focal line, radiating an electric field parallel with the line, suitable for such layered bytes. The desired function of lens is indeed observed in Fig. 3c. We should mention that in these designs, no optimization has been performed, and therefore the reflection from the lens front face is present. Here, our interest is to demonstrate the versatility of the notion of digital metamaterials in designing various structures with spatial variation in their effective permittivity, without resorting to any extensive optimization. Needless to say, one can always add more level of sophistication in the design at a later stage.

*Digital hyperlens*: Another example with special functionality, to which we apply the notion of digital metamaterial, is the 2D hyperlens [9–11], in which circularly concentric pairs of thin layers of positive-permittivity and negative-permittivity materials are packed together in order to "guide" the field radiated by point sources located at the innermost circular layer, along straight lines in the radial direction towards the outside of the lens, thus creating images of the original sources, at the outermost layer where the relative distance between these images is greater than the original distance between the sources. These hyperlenses allow magnification of the image of a sample, before being observed by conventional diffraction-limited imaging optics. In such a device, the values of relative permittivity of each layer, as well as the thickness of the layers, control the dispersion within the lens. We select this example in order to show that even in the very near-field region of our digital bytes (since the layers are all adjacent and tightly packed) these units behave as expected and indeed provide, with good accuracy, the required effective permittivity we need for such hyperlens. In this example, the entire digital hyperlens is designed to be made of the same two material bits, i.e., Ag and $SiO_2$, as mentioned in the beginning of the section. The positive-permittivity layers are entirely made of SiO$_2$, while the negative-permittivity layers are assumed to be made of our core-shell digital bytes, which themselves are made of SiO$_2$ for core and Ag for shell. In fact, due to the hyperlens requirements, once the permittivity of the positive layer is selected (and in this case it corresponds to the permittivity of SiO$_2$ $\varepsilon_d = 2.42$), the permittivity of the negative layer has to be determined following the required dispersion needed to achieve the hyperlensing effect[9,10]. The required negative effective permittivity is then found to have a real part of -2.16, which belongs to the interval $[e_m, e_d]$, and thus for our material bits in the core-shell bytes we should have SiO$_2$ as core and Ag as shell, with a radii ratio of 0.367 (Fig. 3d). The lens, which has inner radius of $l_0/4$ and outer radius of $5l_0/4$, is illuminated by a short electric dipole, laying in the *x-y* plane and parallel with, while separated $l_0/200$ from, the first silica innermost layer, as shown in Fig. 3d. Furthermore, all the layers have the same thickness of $l_0/20$. The results of our simulations are shown in the right panel of Fig. 3d as the field map of the distribution of the amplitude of the electric field vector in this digital lens. The left panel shows the simulation results for the corresponding equivalent hyperlens in which the core-shell bytes were replaced by homogeneous layers with effective permittivity of the digital byte, i.e., $-2.16 + i0.38$. We notice that although the digital bytes operate in very close proximity of each other, the field guidance through the digital hyperlens and the corresponding conventional hyperlens lens is of the same level and of the similar distributions, thus confirming that the digital hyperlens functions as expected using the combination of only two material bits.

*Digital ENZ-based supercoupling*: The final example deals with the use of digital metamaterials in demonstrating the supercoupling phenomenon present in the epsilon-near-zero (ENZ)-filled narrow channels connecting two 2D parallel-plate metallic waveguides. It is known[5,6] that when two identical parallel-plate 2D waveguides with arbitrary orientations with respect to each other are connected via a highly narrow channel, the incoming transverse electromagnetic (TEM) wave from the first waveguide can be tunneled through to the output waveguide when the connecting channel is filled with the ENZ materials regardless of the length of the channel (Fig. 4a). Here, we numerically examine this phenomenon using the digital metamaterial imitating the filling ENZ material. We use the core-shell digital bytes to construct the effective ENZ material for this purpose. Figure 4a shows the schematic of the narrow connecting channel with perfectly electric conducting (PEC) walls, filled with the core-shell digital bytes properly designed to provide the near zero effective relative permittivity. Our simulation results for the TEM wave propagation through the input waveguide, tunneling through the core-shell-filled channel, and reaching the output waveguide, with two different channel lengths are shown in Fig. 4b (in this series of simulation, we consider our two materials bits to be $SiO_2$ and Ag with no loss, i.e., $\varepsilon_d = 2.42$ and $\varepsilon_m = -4.70$, since the phenomenon of supercoupling is most pronounced when the loss is kept at zero). The results reveal the tunneling occurrence at the design frequency of 740.23 THz, where our digital metamaterial exhibit near-zero relative permittivity. A higher second frequency with total transmission is also present. However, this second tunneling is due to the Fabry-Perot effect, which depends on the length of the channel. This is clearly seen by examining and comparing the simulation results for the two cases with two different lengths shown in Fig. 4b (where the channel lengths are $0.45\lambda_0$ and $0.35\lambda_0$). In both cases, the supercoupling tunneling occurs at the same frequencies, whereas the second Fabry-Perot-type tunneling happens at two different frequencies due to two different channel lengths. We also examine numerically the robustness of such tunneling for our ENZ digital metamaterials by considering a 120° bent angle of the port waveguides (shown in Fig. 4c). No substantial difference is observed for the frequency of the ENZ tunneling for this scenario using the same ENZ digital metamaterials.

In conclusion, we have introduced the concept of digital metamaterials, demonstrating that various structures with a variety of requirements for permittivity distributions can be designed using only two material bits. The choice of these bits is rather flexible as long as they possess permittivity with oppositely signed real parts in the frequency band of interest. This brings tremendous simplicity to design of complex metamaterials, and opens doors to novel designs for material bytes as building blocks for a variety of applications in the wide range of electromagnetic spectrum.

## METHODS SUMMARY

**Scattering from digital cylinders.** The simulations of the scattering from both layered cylinders and core-shell cylinders were all performed in COMSOL Multiphysics[®29] using the frequency-domain solver and a tetrahedral mesh. For the simulation of the layered cylinders of Fig. 2c and Fig. 2d, the infinitely extent cylinder has been simulated by replicating the basic unit cell (silica-silver pair) with either PEC or PMC boundary conditions, depending on the polarization of the incoming plane wave. Regarding the core-shell cylinders, the infinite cylinders were simulated by using their equivalent 2D problem.

**Lenses.** The digital dielectric lens was formed in MATLAB® and simulated in COMSOL Multiphysics®[29] as a 2D problem, using the LiveLink™ between the two programs. In the simulation, the lens was illuminated by a plane wave and the amplitude of the electric field vector was computed. The spatial variation of permittivity required for the graded-index lenses was computed using a ray-optics approximation at the central location of each byte. While the graded-index lens made of core-shell cylinders was simulated in a 2D environment in COMSOL Multiphysics®[29], the version with planar bytes was simulated in CST Microwave Studio[30] and the infinite extent along the direction normal to the plane of the lens was taken into account by imposing perfect magnetic conductor boundary conditions. Eventually, both the homogeneous and the digital version of the hyperlens were simulated in CST Microwave Studio[31] following a similar approach as for the planar-byte graded-index lens.

**ENZ–based supercoupling.** The supercoupling through the ENZ channel was simulated in a 2D environment using COMSOL Multiphysics®[29]. The power flux lines reported in Fig. 4a and in the Supplementary Information are the lines resulting from the simulations.

**Acknowledgement**

This work is supported in part by the US Office of Naval Research (ONR) Multidisciplinary University Research Initiative (MURI) Program grant number N00014-10-1-0942.

**References**

1.      Sessions, R. How a "difficult" composer gets that way. *New York Times* 89 (1950).

2.      A Letter From The Publisher: Dec. 14, 1962. *Time* (1962).

3.      Boole, G. *An Investigation of the Laws of Thought on Which are Founded the Mathematical Theories of Logic and Probabilities*. (Macmillian, 1854).

4.      Shannon, C. E. A symbolic analysis of relay and switching circuits. *Am. Inst. Electr. Eng. Trans.* **57,** 713–723 (1938).

5.      Silveirinha, M. & Engheta, N. Tunneling of Electromagnetic Energy through Subwavelength Channels and Bends using ε-Near-Zero Materials. *Phys. Rev. Lett.* **97,** 157403 (2006).

6.      Silveirinha, M. & Engheta, N. Theory of supercoupling, squeezing wave energy, and field confinement in narrow channels and tight bends using ε near-zero metamaterials. *Phys. Rev. B* **76,** 245109 (2007).

7.      Alù, A., Silveirinha, M. & Engheta, N. Transmission-line analysis of ε-near-zero–filled narrow channels. *Phys. Rev. E* **78,** 016604 (2008).

8.      Edwards, B., Alù, A., Silveirinha, M. & Engheta, N. Experimental Verification of Plasmonic Cloaking at Microwave Frequencies with Metamaterials. *Phys. Rev. Lett.* **103,** 153901 (2009).

9.      Salandrino, A. & Engheta, N. Far-field subdiffraction optical microscopy using metamaterial crystals: Theory and simulations. *Phys. Rev. B* **74,** 075103 (2006).

10.     Jacob, Z., Alekseyev, L. V & Narimanov, E. Optical Hyperlens: Far-field imaging beyond the diffraction limit. *Opt. Express* **14,** 8247–56 (2006).

11.     Liu, Z., Lee, H., Xiong, Y., Sun, C. & Zhang, X. Far-field optical hyperlens magnifying sub-diffraction-limited objects. *Science (80-. ).* **315,** (2007).

12.     Engheta, N. & Ziolkowski, R. *Electromagnetic Metamaterials: Physics and Engineering Explorations*. 440 (Wiley-IEEE Press, 2006).

13.     Eleftheriades, G. V. & Balmain, K. G. *Negative Refraction Metamaterials: Fundamental Principles and Applications*. 440 (Wiley-IEEE Press, 2005).

14.     Smith, D. R., Pendry, J. B. & Wiltshire, M. C. K. Metamaterials and negative refractive index. *Science* **305,** 788–92 (2004).

15.     Caloz, C. & Itoh, T. *Electromagnetic Metamaterials: Transmission Line Theory and Microwave Applications*. 376 (Wiley-IEEE Press, 2013).

16.     Cai, W. & Shalaev, V. M. *Optical Metamaterials: Fundamentals and Applications*. 212 (Springer, 2009).

17.     Shevchenko, E. V, Talapin, D. V, Kotov, N. A., O'Brien, S. & Murray, C. B. Structural diversity in binary nanoparticle superlattices. *Nature* **439,** 55–9 (2006).

18.    Liu, N., Liu, H., Zhu, S. & Giessen, H. Stereometamaterials. *Nat. Photonics* **3,** 157–162 (2009).

19.    Sun, Y., Edwards, B., Alù, A. & Engheta, N. Experimental realization of optical lumped nanocircuits at infrared wavelengths. *Nat. Mater.* **11,** 208–12 (2012).

20.    Monticone, F., Estakhri, N. M. & Alù, A. Full Control of Nanoscale Optical Transmission with a Composite Metascreen. *Phys. Rev. Lett.* **110,** 203903 (2013).

21.    Caglayan, H., Hong, S.-H., Edwards, B., Kagan, C. R. & Engheta, N. Near-Infrared Metatronic Nanocircuits by Design. *Phys. Rev. Lett.* **111,** 073904 (2013).

22.    Milton, G. W. Bounds on the transport and optical properties of a two-component composite material. *J. Appl. Phys.* **52,** 5294 (1981).

23.    Soukoulis, C. M. & Wegener, M. Optical metamaterials--More Bulky and Less Lossy. *Science* **330,** 1633–4 (2010).

24.    Gansel, J. K. *et al.* Gold helix photonic metamaterial as broadband circular polarizer. *Science* **325,** 1513–5 (2009).

25.    Sihvola, A. H. *Electromagnetic Mixing Formulas and Applications*. 284 (IET, 1999).

26.    Chettiar, U. K. & Engheta, N. Internal homogenization: effective permittivity of a coated sphere. *Opt. Express* **20,** 22976–86 (2012).

27.    Johnson, P. B. & Christy, R. W. Optical Constants of the Noble Metals. *Phys. Rev. B* **6,** 4370–4379 (1972).

28.    Bass, M. at al. *Handbook of Optics*. (McGraw Hill, 1995).

29.    www.comsol.com.

30.    www.cst.com.

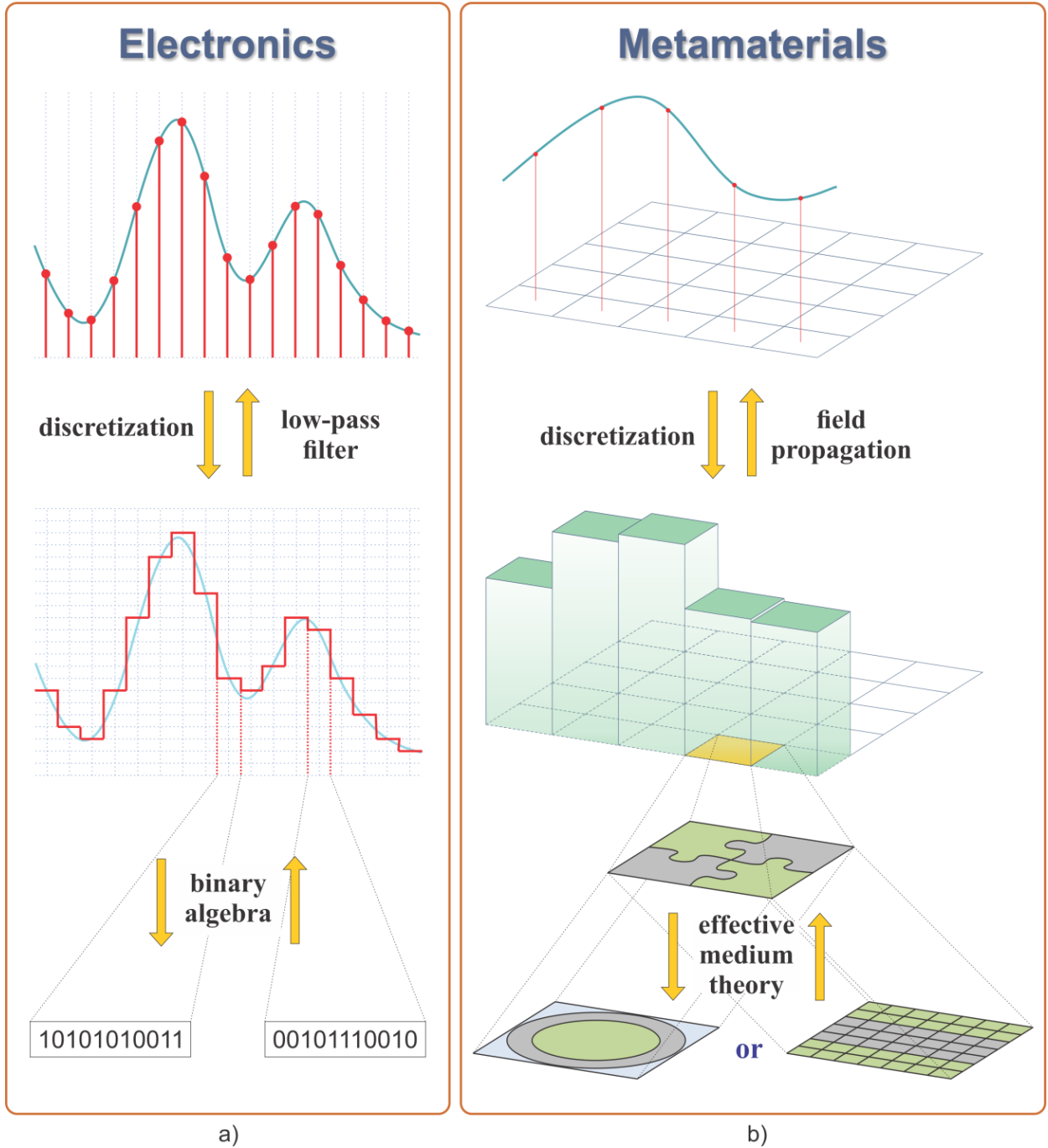

**Figure 1 | Notion of digital metamaterials inspired by digital electronics. a**, analog-to-digital and digital-to-analog signal conversion in electronics as an essential step to use only two digital bits "0" and "1" for representing any number (logic states in the Boolean algebra). **b**, sampling, discretizing, and "digitizing" materials with required permittivity function, and then designing it with two "metamaterial bits" as building blocks for "digital metamaterials"..

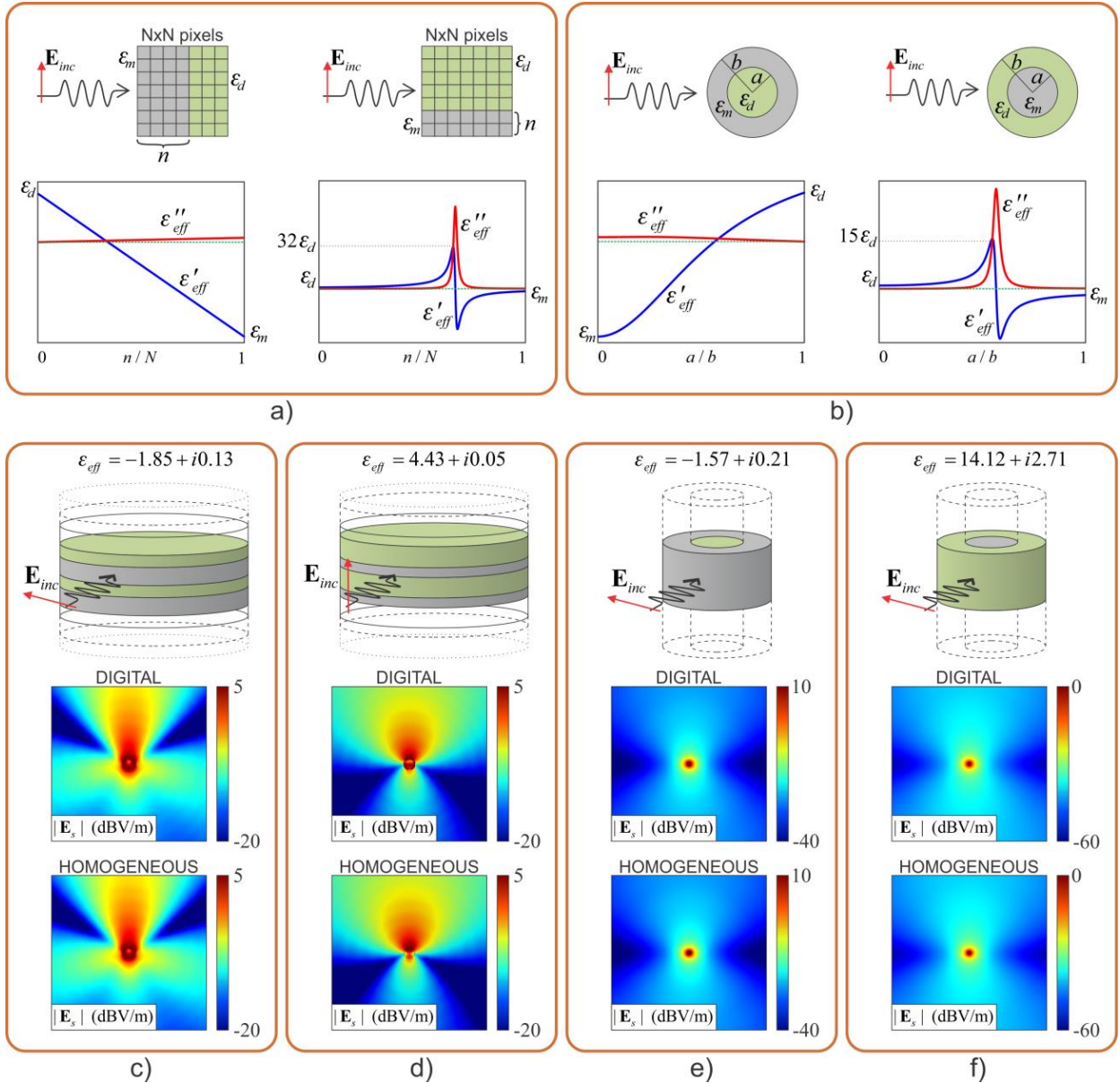

**Figure 2 | Scattering from the digital metamaterial "bytes". a**, effective permittivity $\varepsilon_{eff} = \varepsilon'_{eff} + i\varepsilon''_{eff}$ of a two-dimensional (2D) planar digital byte for two different polarizations of the incident electric field, vs. the volume fraction $n/N$, and **b**, effective permittivity of 2D core-shell cylindrical bytes with two different arrangements of material 'bits', vs. the ratio of radii $a/b$. The "bits" have relative permittivity $\varepsilon_m = -4.7 + i0.22$ (silver (gray)) and $\varepsilon_d = 2.42$ (SiO$_2$ (green)). **c** and **d**, comparison of scattering from an infinite cylinder composed of thin planar bytes for incident electric field parallel with (**c**) and perpendicular to (**d**) the layers of bytes, and the equivalent homogeneous cylinders with effective permittivity. Simulation results show the distributions of the magnitude of the scattered electric field vector (V/m, in dB, with amplitude of the incident electric field of 1 V/m, denoted as 0 dBV/m). **e** and **f**, similar to (**c**) and (**d**), but for the core-shell cylindrical cells designed with silica for core and silver for shell (**e**) and with silver for core and silica for shell (**f**). Similarity in scattering signatures of each metamaterial byte with its homogenous counterpart is clearly noticeable.

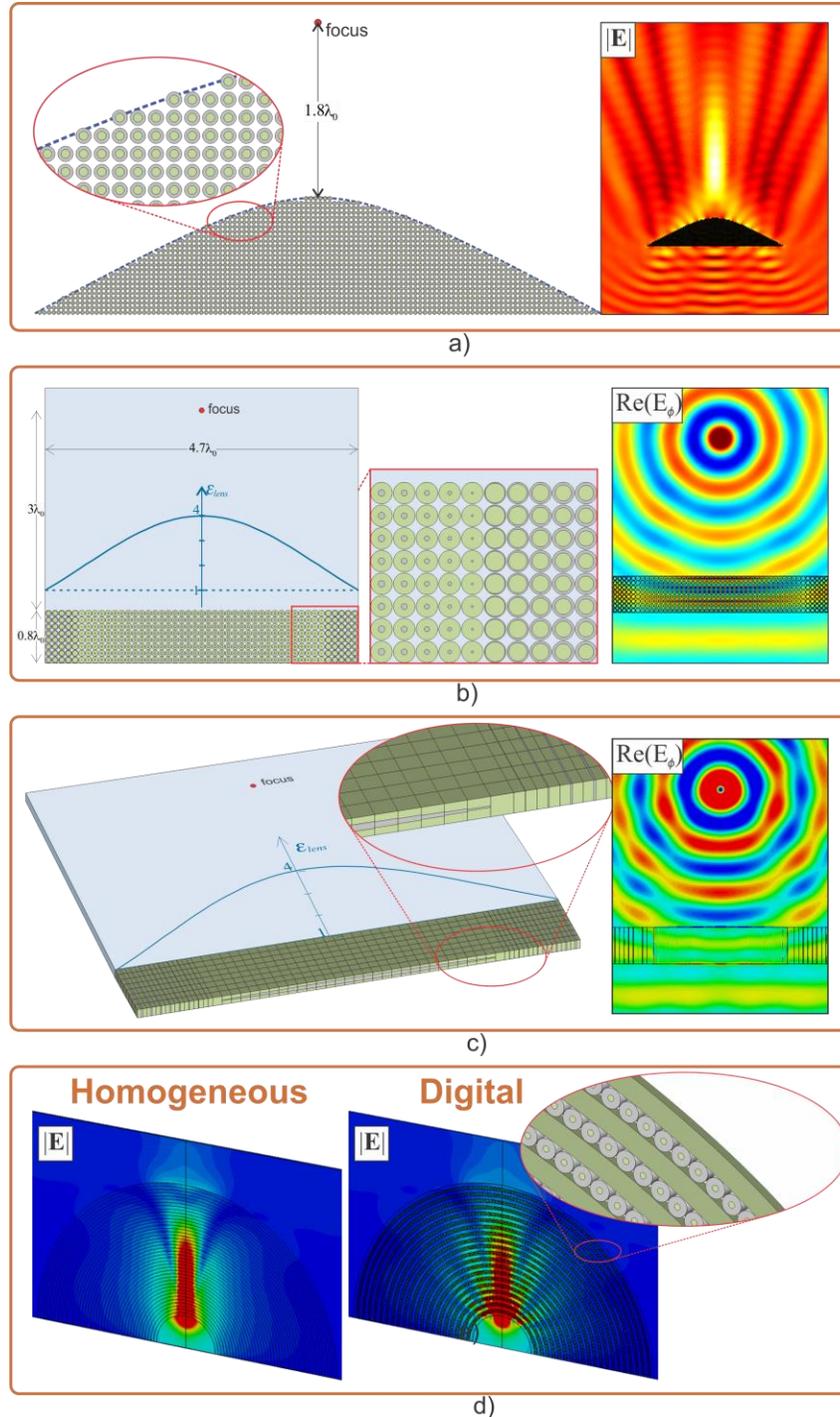

a)

b)

c)

d)

**Figure 3 | Lenses designed with digital metamaterials. a**, convex lens with hyperbolic profile. Rectangular flat graded-index lens formed by (**b**) core-shell bytes and (**c**) planar cells. **d**, hyperlens with positive-epsilon layers designed with silica, and negative-epsilon layers designed with core-shell bytes (also a mix of silica and silver). The relative permittivity values for silver and SiO₂ used in these simulations are given in Fig. 2.

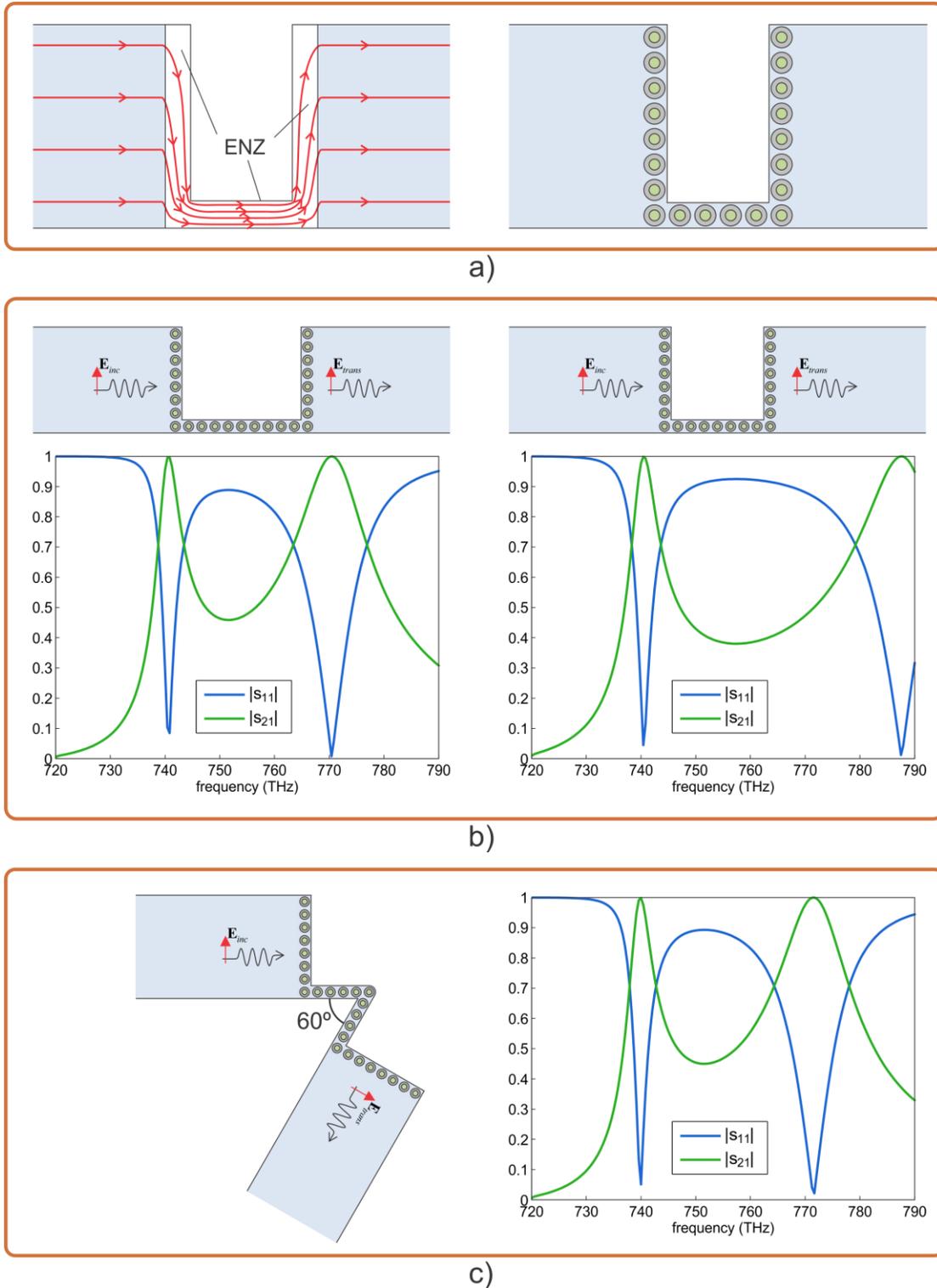

**Figure 4 | ENZ tunneling using digital metamaterials. a**, geometry and sketch of power flow tunneling through a narrow channel filled with a homogeneous ENZ (left) and its design using digital core-shell elements (right). **b**, Simulation results for the scattering parameters (reflection ($s_{11}$) and transmission ($s_{21}$)) through the channel with length $0.45\lambda_0$ (left) and $0.35\lambda_0$ (right). **c**, scattering parameters for a 120° bent channel.

# SUPPLEMENTARY INFORMATION

# Digital Metamaterials

*Cristian Della Giovampaola and Nader Engheta[1]*

*University of Pennsylvania*

*Department of Electrical and Systems Engineering*

*Philadelphia, Pennsylvania 19104, USA*

Email: dcr@seas.upenn.edu, engheta@ee.upenn.edu

## 1. Supplementary results on scattering

The following results are additional findings to the field maps shown in Fig. 2c through Fig. 2f. For each of the structures of Fig. 2, the amplitude of the scattered electric field vector produced by the digital version of each cylinder (either made of layered bytes or core-shell bytes) is compared to those provided by its equivalent homogeneous version along circular scans at different distances from the cylinder.

**Planar-byte cylinder.**

These results refer to the structure depicted in Fig. 2c and Fig. 2d in the main text. The angular scans are performed along circles with radius ranging from $0.4\lambda_0$ (top left) to $3\lambda_0$ (bottom right). Since the radius of the cylinder is $0.2\lambda_0$, this implies that for the angular scan with the smallest radius, the scattered electric field is probed at $0.2\lambda_0$ from the cylinder lateral surface. The plots of Fig. S1 refer to a TM polarized wave, whereas Fig. S2 reports the field scans relative to a TE illumination.

---

[1] To whom correspondence should be addressed, Email: engheta@ee.upenn.edu

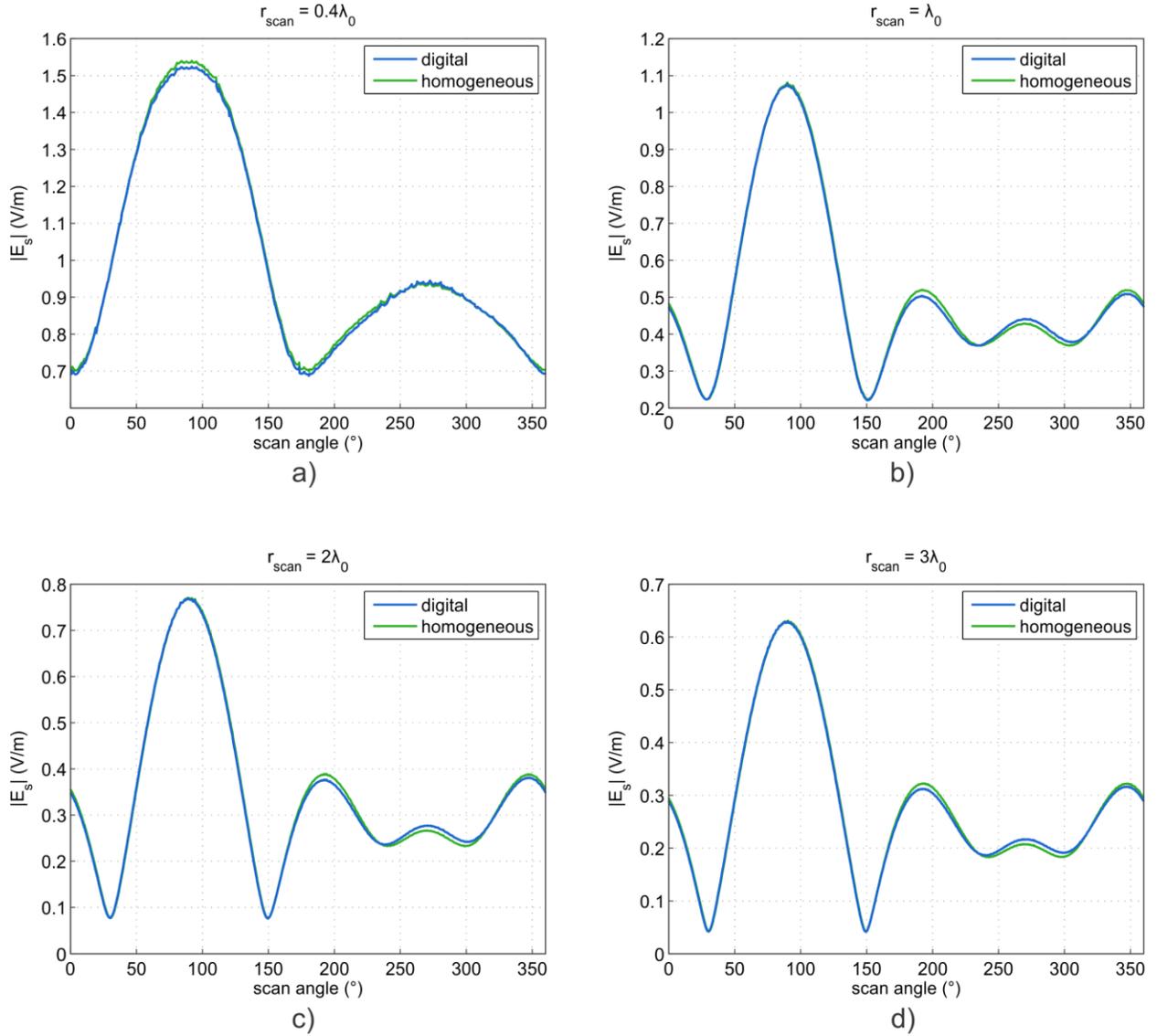

**Figure S1 | Field obtained at circular angular scan for planar-byte cylinder and TM polarization.** Amplitude of the scattered electric field vector for the digital cylinder (blue line) and its equivalent homogeneous version (green line) along circles with radius: **a**, $0.2\lambda_0$; **b**, $\lambda_0$; **c**, $2\lambda_0$; **d**, $3\lambda_0$.

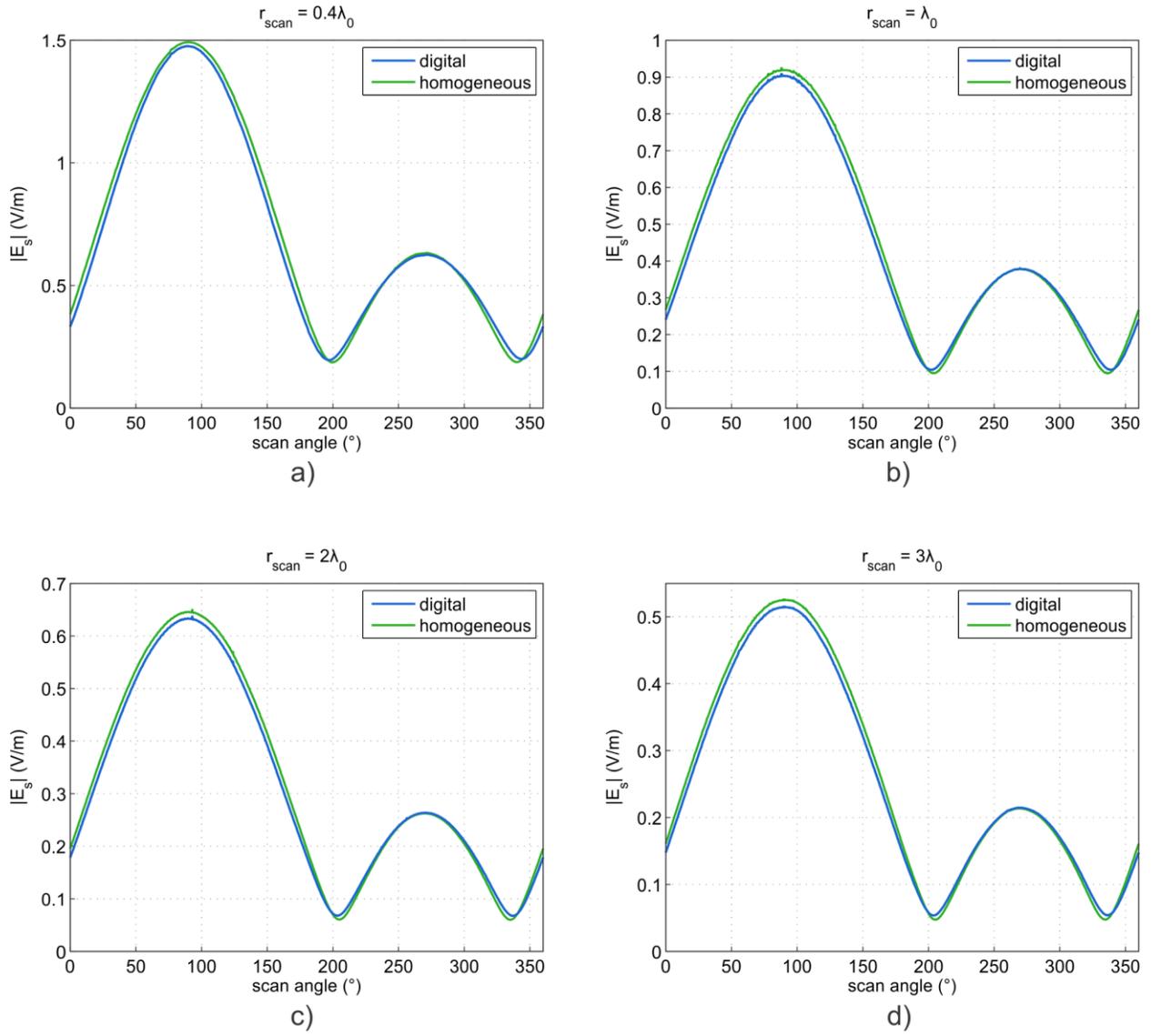

**Figure S2 | Field obtained at circular angular scan for planar-byte cylinder and TE polarization.** Amplitude of the scattered electric field vector for the digital cylinder (blue line) and its equivalent homogeneous version (green line) along circles with radius: **a**, $0.2\lambda_0$; **b**, $\lambda_0$; **c**, $2\lambda_0$; **d**, $3\lambda_0$.

## Core-shell cylinder.

In this section we show the amplitude of the scattered electric field vector on circular angular scans for the core-shell cylinder depicted in Fig. 2e and Fig. 2f of the main text. The analysis is similar to the one previously described about the planar-byte cylinders, but the scan radii now range from $0.075\lambda_0$ to $1.8\lambda_0$.

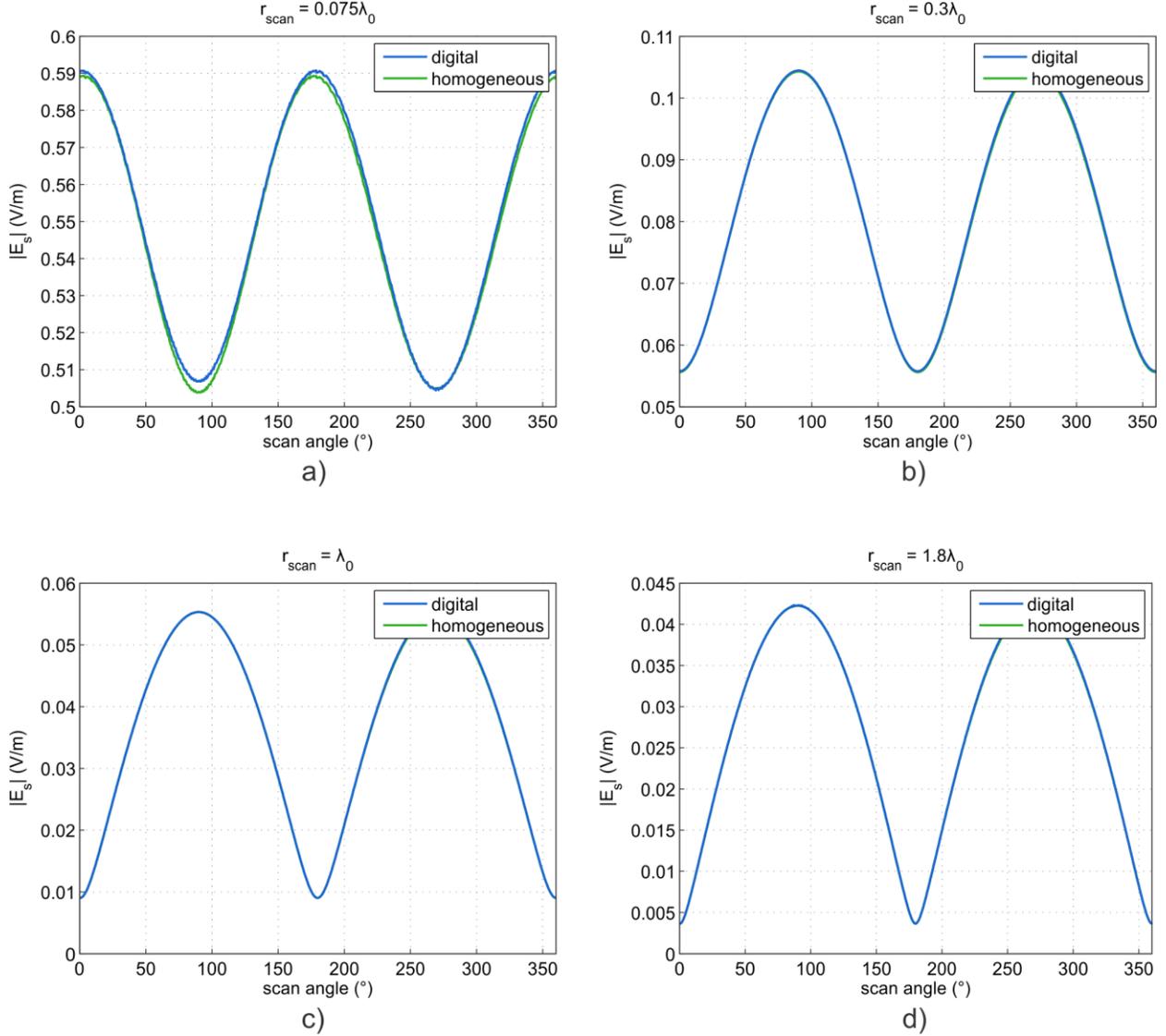

**Figure S3 | Field obtained at circular angular scan for core-shell cylinder with silica-core and silver-shell.** Amplitude of the scattered electric field vector for the digital cylinder (blue line) and its equivalent homogeneous version (green line) along circles with radius: **a**, $0.075\lambda_0$; **b**, $0.3\lambda_0$; **c**, $\lambda_0$; **d**, $1.8\lambda_0$.

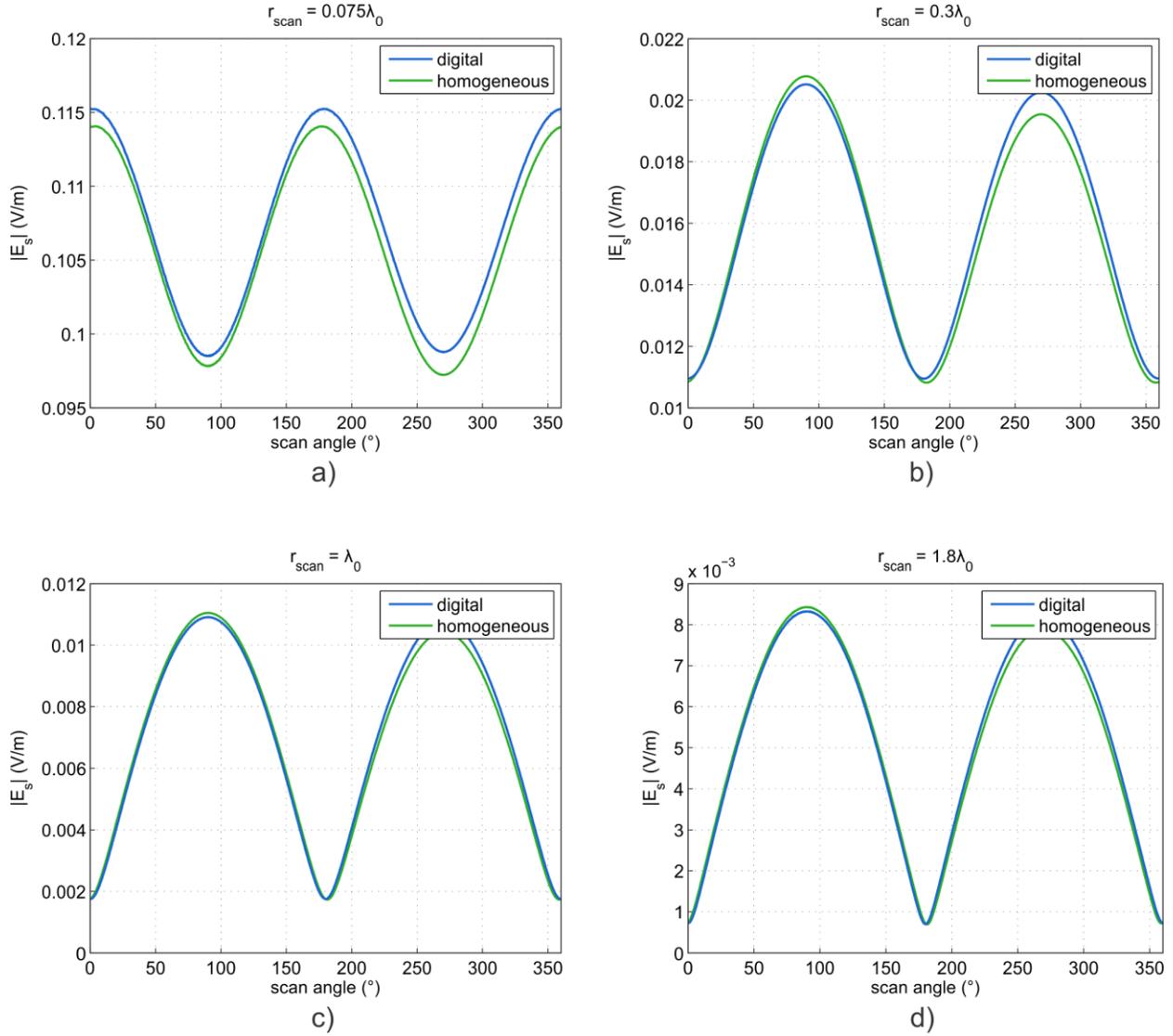

**Figure S4 | Field obtained at circular angular scan for core-shell cylinder with silver-core and silica-shell.**
Amplitude of the scattered electric field vector for the digital cylinder (blue line) and its equivalent homogeneous version (green line) along circles with radius: **a**, $0.075\lambda_0$; **b**, $0.3\lambda_0$; **c**, $\lambda_0$; **d**, $1.8\lambda_0$.

## 2. Material Loss in the convex lens.

The effect of the loss in the silver core of the core-shell bytes has been evaluated comparing the electric field vector amplitude of the digital convex lens with the one produced by an equivalent homogeneous lens where the real and imaginary part of the relative permittivity equal the real and imaginary part of the effective permittivity of the core-shell byte evaluated through numerical retrieval. As we can see from Fig. S5, the effects of the loss are more important for the case of a homogeneous lossy lens, whereas in the digital lens the losses play a less crucial role.

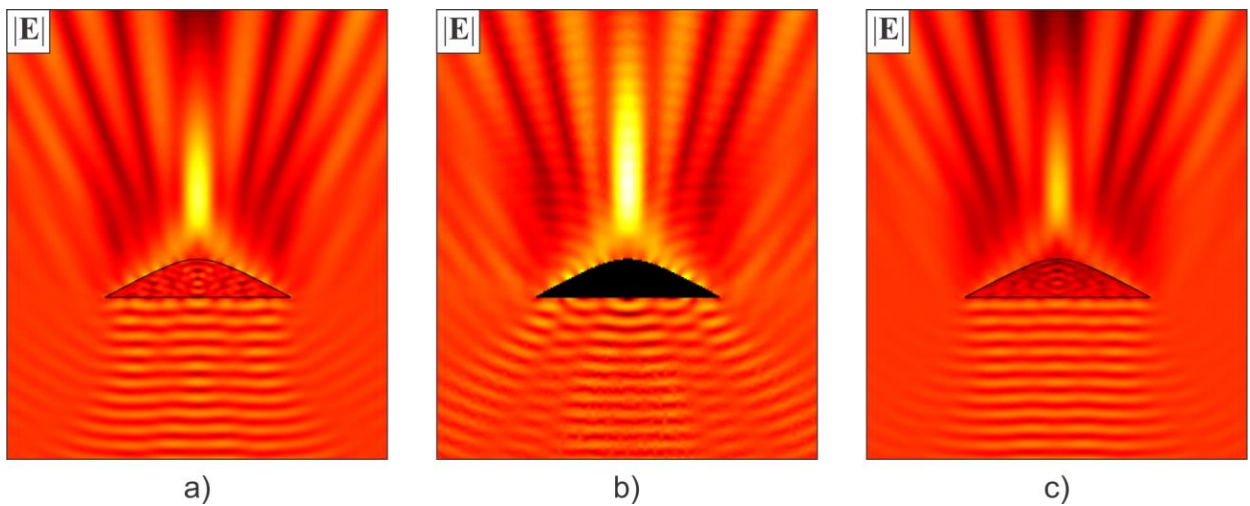

a)                              b)                              c)

**Figure S5 | Effect of the material loss for the convex lens.** Amplitude of the electric field vector. **a**, homogeneous lossless convex lens with real relative permittivity of 4. **b**, digital version of the lens with core-shell cylinders. **c**, homogeneous version where the relative permittivity of the lens has the same real and imaginary part of the core-shell byte (from the parameter retrieval). In this case the effect of the loss of the silver in the total field is limited and it seems not to be as crucial as if the loss were equally distributed all over the lens.

## 3. Graded-index lens dispersion

The graded-index lenses shown in Fig. 3b and Fig. 3c in the main text have been designed using a ray optics approximation by imposing equal electrical paths from the focal point to the output plane. The dots in the blue curve in Fig. S6 represent the value of required permittivity of each discrete strip of the lens (whose width corresponds to the byte size, which for the current design was $0.1\lambda_0$ for both lenses of Fig. 3b and Fig. 3c). In designing the lens permittivity, the value of permittivity at the sides of the lens was chosen to be 1 in order to match the surrounding air (although this is not a general requirement). Because of the stated constraint and the lens dimensions, the central value for the permittivity of the lens turns to be 4. In Fig. S6, the lens permittivity is also compared with the values of permittivity of silica and silver (in this case, the real part of it). As mentioned in the main text, the above design has not been optimized to make the reflections minimum, although the digital metamaterial approach can be used for more sophisticated designs.

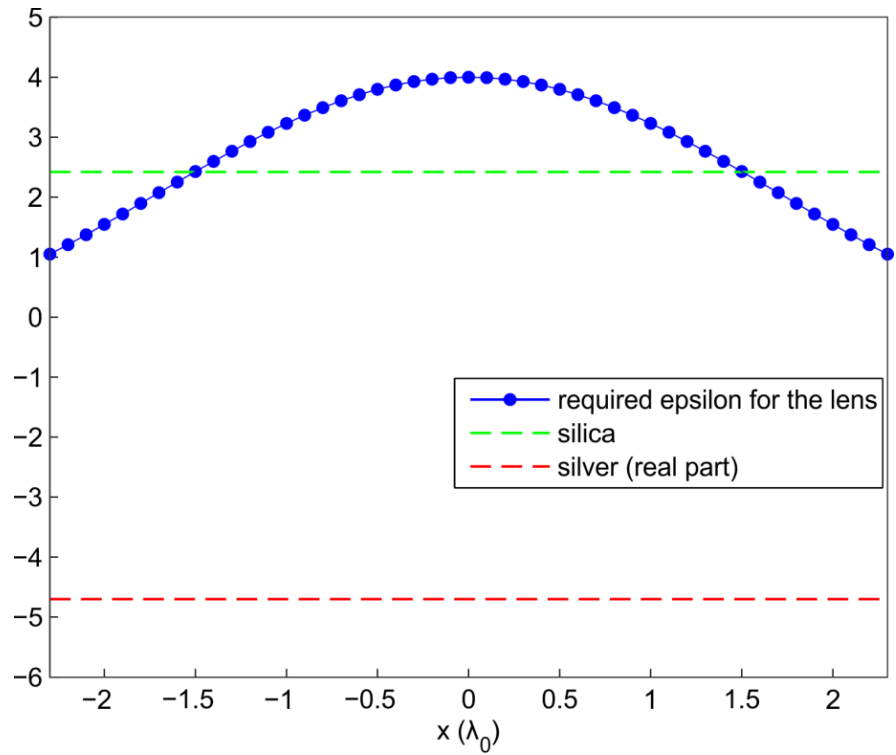

**Figure S6 | Relative permittivity distribution required for this graded-index lens.** The spatial distribution of the relative permittivity that is required to achieve the lensing effect is compared with the permittivity of the silica (green dashed line) and with the real part of the permittivity of the silver (red dashed line).

### 4. Graded-index lens at 1.55 μm

In this additional study, which for the sake of brevity was not reported in the main body of the paper, we describe an alternative design of the digital graded-index lens. The lens is made of core-shell bytes and its structure is similar to the one presented in Fig. 3b. However, the materials used as metamaterial bits and the operating frequency are different. Here we show a possible implementation for applications in the telecommunication wavelength, specifically 1.55 μm and also we want to demonstrate the generality of our concept of digital metamaterials by highlighting that its use is not constrained by specific materials or frequencies. In this example, silicon (Si) was chosen as the material byte with positive permittivity ($\varepsilon_{Si}$ = 12.09, as reported in the main text with the related reference) due to its easy integration with fabrication techniques and its wide use for telecommunication. Concerning the medium with negative permittivity, indium tin oxide (ITO) was selected for its moderately negative value of real permittivity and relatively low loss ($\varepsilon_{ITO}$ = -1.1+$i$0.32, obtained from the reference cited in the main text) at the wavelength of 1.55 μm. The dimensions of the lens are exactly the same as the ones reported in Fig. 3b and Fig. 3c. We notice that the field map depicted in Fig. S7 strongly resembles the ones of Fig. 3b and Fig. 3c, albeit at a different operating wavelength using different materials as metamaterials bits.

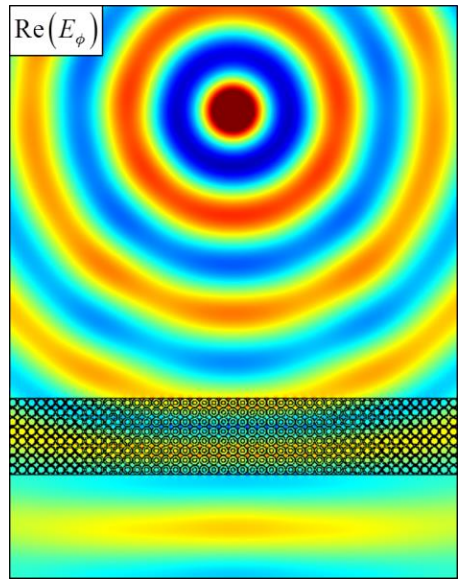

**Figure S7 | Field wavefronts for the graded-index lens at 1.55 μm.** Field map showing the evolution of the $\phi$-component of the electric field in a digital graded-index lens where the two material bits are made of silicon and indium tin oxide (ITO).

## 5. Power flow in the digital ENZ-based supercoupling.

The results on the digital ENZ-based supercoupling reported in Fig. 4 show the transmission characteristics of a narrow digital ENZ channel obtained using core-shell bytes embedded in free space, so that the effective epsilon of each byte would be vanishing at the working frequency. In this additional study, reported in Fig. S8 and Fig. S9, we analyzed the power flux lines inside the ENZ channel for both the homogeneous and the digital case. In fact, by comparing the power flow lines of both the structures, we notice strong similarities in the way the energy is guided and "squeezed" into the channel, although the presence of the core-shell particles locally affects the distribution of the power flux lines. These additional results confirm that the supercoupling effect is due to the vanishing effective permittivity of the channel between the two waveguides and not to some other unforeseen phenomenon.

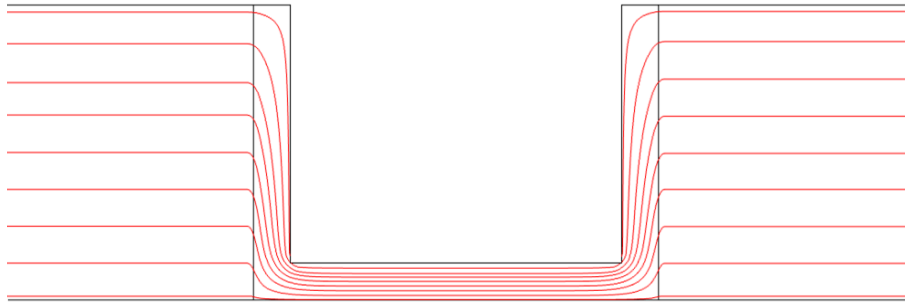

**Figure S8 | Power flow in the ideal homogeneous ENZ supercoupling.** Power flow lines across the narrow homogeneous ENZ channel responsible for the supercoupling showing how the energy is "squeezed" into the channel.

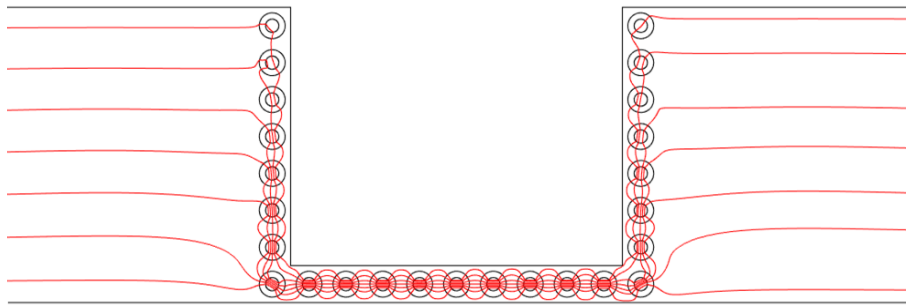

**Figure S9 | Power flow in the digital ENZ supercoupling.** Power flow lines through the digital ENZ channel formed by metamaterial bytes.